\documentclass[11pt]{article}
\usepackage[left=1in,top=1in,right=1in,bottom=1in,head=.1in,nofoot]{geometry}

\usepackage{setspace,url,bm,amsmath} % For double-spacing, URL font, math symbols

\usepackage{titlesec} % Section header formatting
\titlelabel{\thetitle.\quad} % Section header formatting
\usepackage{graphicx} % Graphics scaling
\usepackage{bbm}
\usepackage{latexsym}
\usepackage{caption}
\usepackage[margin=20pt]{subcaption}
\usepackage{hyperref}
\usepackage{enumerate}

\usepackage[table]{xcolor}

\newcommand{\GG}[1]{}

\usepackage{amsthm}
\usepackage{comment}
\theoremstyle{definition}

\newtheorem*{theorem*}{Theorem}

\newtheorem*{corollary*}{Corollary}

\usepackage{natbib} % ASA citation style
\bibpunct{(}{)}{;}{a}{}{,} % ASA citation style

\usepackage{etoolbox} % Bibliography underfull/overfull box fix
\apptocmd{\sloppy}{\hbadness 10000\relax}{}{} % Bibliography underfull/overfull box fix

\usepackage{color}
\usepackage{listings}

\def\var{\text{var}}
\def\cov{\text{cov}}
\def\sumn{\sum_{i=1}^n}

\begin{document}
\doublespacing
\title{\bf \Large Two seemingly paradoxical results in linear models:\\ 
the variance inflation factor and the analysis of covariance}
\author{Peng Ding\footnote{
Peng Ding (Email: \url{pengdingpku@berkeley.edu}) is Assistant Professor in the Department of Statistics, University of California, Berkeley, CA 94720, USA. 
}
}
\date{}
\maketitle

\begin{abstract}
A result from a standard linear model course is that the variance  of the ordinary least squares (OLS) coefficient of a variable will never decrease when including additional covariates. The variance inflation factor (VIF) measures the increase of the variance. Another result from a standard linear model or experimental design course  is that including additional covariates in a linear model of the outcome on the treatment indicator will never increase the variance of the OLS coefficient of the treatment at least asymptotically. This technique is called the analysis of covariance (ANCOVA), which is often used to improve the efficiency of treatment effect estimation. So we have two paradoxical results: adding covariates never decreases the variance in the first result but never increases the variance in the second result. In fact, these two results are derived under different assumptions. More precisely, the VIF result conditions on the treatment indicators but the ANCOVA result averages over them. Comparing the estimators with and without adjusting for additional covariates in a completely randomized experiment, I  show  that the former has smaller variance averaging over the treatment indicators, and the latter has smaller variance at the cost of a larger bias conditioning on the treatment indicators. Therefore, there is no real paradox. 

\end{abstract}

\section{Variance inflation factor}\label{sec::vif}

Consider the following linear regression:
\begin{eqnarray}
\label{eq::lm}
y_i = \alpha +  \tau z_i + \beta ' x_i + \varepsilon_i,\quad (i=1,\ldots, n)
\end{eqnarray}
where $z_i$ is a scalar and $x_i$ is a scalar or vector. Without loss of generality, we assume $\bar{x} = n^{-1} \sum_{i=1}^n x_i =0$. 
In a leading example, $z_i$ is the treatment variable and $x_i$ contains all the covariates. Using a standard result in linear models, we can write the OLS estimator for $\tau$ as 
$$
\hat{\tau}_\text{a} = \frac{   \sumn \check{z}_i y_i   }{  \sumn \check{z}_i ^2 },
$$ 
where $\check{z}_i $ is the residual from the OLS fit of $z_i$ on $(1, x_i)$. This result is also called
the Frisch--Waugh--Lovell Theorem in econometrics. 
If the regressors $(z_i,x_i)$'s are all fixed and the $\varepsilon_i$'s are independent and identically distributed (IID) with mean $0$ and variance $\sigma^2$ as in the classic linear model, then   the variance of $\hat{\tau}_\text{a}$ equals 
\begin{eqnarray}
\label{eq::var}
\var(\hat{\tau}_\text{a}) &=& \frac{   \sumn \check{z}_i^2 \var(y_i)   }{ \left(  \sumn \check{z}_i ^2 \right)^2 } \nonumber \\
&=& \frac{\sigma^2}{  \sumn \check{z}_i ^2 } \nonumber   \\
&=& \frac{\sigma^2}{  \sumn (z_i - \bar{z}) ^2 } \times \frac{   \sumn (z_i - \bar{z} )^2  }{  \sumn \check{z}_i ^2 } .
\end{eqnarray}
The first term of \eqref{eq::var} equals the variance of
\begin{eqnarray}
\label{eq::unadjust}
\hat{\tau} = \frac{   \sumn (z_i - \bar{z}) y_i   }{  \sumn (z_i - \bar{z}) ^2 } ,
\end{eqnarray}
i.e., the coefficient of $z_i$ in the OLS  fit of $y_i$ on $(1 , z_i )$ without adjusting for $x_i$. The second term of \eqref{eq::var} is the VIF, which is no smaller than 1 because it is the total sum of squares divided by the residual sum of squares in the OLS fit of $z_i$ on $(1, x_i).$  The VIF can be equivalently written as $(1-R^2_{z\mid x})^{-1}$, where $R^2_{z\mid x}$ is the sample $R^2$ between $z_i$ and $x_i$. So the VIF result can also be written as
$$
\var(\hat{\tau}_\text{a}) = \var(\hat{\tau}) \times (1-R^2_{z\mid x})^{-1}.
$$
It highlights the bias-variance tradeoff: with more covariates $x_i$ included, the model is closer to the truth and thus leads to smaller bias in estimating $\tau$, but at the same time it results in larger variance of $\hat{\tau}_\text{a} $. 
See \citet{faraway2016linear}, \citet{fox2015applied} and \citet{agresti2015foundations} for textbook discussions. 

Thus, from \eqref{eq::var}, the variance of $\var(\hat{\tau}_\text{a})$ will never decrease with more covariates in \eqref{eq::lm}, because the residual sum of squares $\sumn \check{z}_i ^2$ will decrease while the total sum of squares $\sumn (z_i - \bar{z} )^2 $ is constant.  An immediate result is  
$$
\var(\hat{\tau}_\text{a}) \geq \var(\hat{\tau}) ,
$$ 
and the equality holds when $X'Z=0$, where $Z=(z_1,\ldots, z_n)'$ is the vector formed by regressors $z_i$'s and $X=(x_1',\ldots,x_n')'$ is the  matrix formed by the regressors $x_i$'s. The orthogonality of regressors (i.e., $X'Z=0$) ensures that $R^2_{z\mid x} =0$ and $\hat{\tau} = \hat{\tau}_\text{a}$.

\section{Analysis of covariance}\label{sec::ancova}

Now we consider a special case of \eqref{eq::lm}: the $x_i$'s are pretreatment covariates, the $z_i$'s are the binary treatment indicators ($1$ for the treatment and $0$ for the control), and the $y_i$'s are the outcomes of interest. Then \eqref{eq::lm} is the standard ANCOVA model, and the parameter of interest $\tau$ is the treatment effect. Let $n_1= \sumn z_i$ and $n_0 = \sumn (1-z_i)$ be the numbers of units under the treatment and control, respectively. 
As in Section \ref{sec::vif}, we assume that the $\varepsilon_i$'s are IID with mean $0$ and variance $\sigma^2$. In
a completely randomized experiment, we  further assume that the $z_i$'s are from random permutation of $n_1$ $1$'s and $n_0$ $0$'s.

Because $z_i$ is binary, we can define 
$$
\hat{\delta}_x = n_1^{-1}\sumn z_i x_i - n_0^{-1} \sumn (1-z_i) x_i ,\quad 
\hat{\delta}_\varepsilon =  n_1^{-1}\sumn z_i \varepsilon_i - n_0^{-1} \sumn (1-z_i)  \varepsilon_i  
$$
as the differences-in-means of $x$ and $\varepsilon$. Further let $\hat{\beta}$ be the OLS estimator for $\beta$ in \eqref{eq::lm}. 
The estimator $\hat{\tau} $ in \eqref{eq::unadjust} without adjusting for covariates simplifies to the difference in means of the outcome:
$$
\hat{\tau} =  n_1^{-1}\sumn z_i y_i - n_0^{-1} \sumn (1-z_i) y_i ,
$$
which further simplifies to
\begin{eqnarray} 
\hat{\tau} &=&  n_1^{-1}\sumn z_i  (\alpha +  \tau z_i + \beta ' x_i + \varepsilon_i)- n_0^{-1} \sumn (1-z_i) (\alpha +  \tau z_i + \beta ' x_i + \varepsilon_i) \nonumber  \\
&=& \tau + \beta' \hat{\delta}_x +   \hat{\delta}_\varepsilon,   \label{eq::adj-2}
\end{eqnarray} 
under \eqref{eq::lm}. Based on OLS of \eqref{eq::lm}, the estimator $\hat{\tau}_\text{a} $ adjusting for the covariates simplifies to 
\begin{eqnarray*}
\hat{\tau}_\text{a}   &  =&  n_1^{-1}\sumn z_i (y_i - \hat{\beta}'x_i) - n_0^{-1} \sumn (1-z_i) (y_i - \hat{\beta}'x_i)   \\
&=&  \tau +( \beta - \hat{\beta}) ' \hat{\delta}_x +  \hat{\delta}_\varepsilon .
\end{eqnarray*}
With large samples, we can ignore the term $( \beta - \hat{\beta}) ' \hat{\delta}_x$  above to obtain
$$
\hat{\tau}_\text{a}  \approx  \tau +\hat{\delta}_\varepsilon, 
$$
because $( \beta - \hat{\beta}) ' \hat{\delta}_x = O_P(n^{-1})$ is of higher order due to $\hat{\beta}  -  \beta = O_P(n^{-1/2})$ and $\hat{\delta}_x =  O_P(n^{-1/2})$, both justified by central limit theorems under certain moment conditions. See \citet{li2017general} for technical details.

Under complete randomization, we can show that 
$$
E(\hat{\delta}_\varepsilon) = 0 ,\quad 
E(\hat{\delta}_x) = 0 
$$
based on a standard result for the difference-in-means  \citep{Neyman23}, 
\begin{eqnarray}
\label{eq::var-cov}
\var(\hat{\delta}_\varepsilon) = \frac{n}{n_1n_0} \sigma^2,\quad
\var(\hat{\delta}_x) = \frac{n}{n_1n_0} S_x^2,\quad
\end{eqnarray}
where $S_x^2 = (n-1)^{-1}\sumn (x_i-\bar{x}) (x_i-\bar{x})'$ is the finite population covariance of the $x_i$'s \citep{ImbensRubin15},
and moreover, the uncorrelatedness of the two difference-in-means \citep{li2017general}
$$
 \cov(\hat{\delta}_\varepsilon, \hat{\delta}_x) = 0.
$$
% The first variance and the third covariance in \eqref{eq::var-cov} follow from standard variance and covariance calculations by first conditioning on all $z_i$'s, and the second variance in \eqref{eq::var-cov} follows from \citet{Neyman23}'s result on the difference-in-means from a completely randomized experiment \citep[c.f.][]{ImbensRubin15, li2017general}.  
Then $E(\hat{\tau} )  = \tau$ and $E(\hat{\tau}_\text{a}) \approx \tau$, i.e., $\hat{\tau}$ is unbiased and $\hat{\tau}_\text{a}$ is consistent for $\tau$. Their variances satisfies
$$
\var(\hat{\tau} ) - \var(\hat{\tau}_\text{a}) \approx  \var(  \beta' \hat{\delta}_x ) =  \frac{n}{n_1n_0}   \beta ' S_x^2 \beta \geq 0. 
$$
Thus, if $\beta \neq 0$ then ANCOVA improves estimation efficiency, at least asymptotically. See \citet{kempthorne1952design}, \citet{Hinkelmann2007} and \citet{cox2000theory} for textbook discussions.

\section{From conflict to unification} 

\subsection{A unified data generating process}\label{subsec::dgp}
 
From the VIF result, we see that adding more covariates never decreases the variance of an OLS coefficient. In contrast, from the ANCOVA result, we see that adding more covariates never increases the variance of an OLS coefficient at least asymptotically. These two results are both standard in textbooks of linear models or experimental designs. However, they seem to give opposite conclusions. Both results are derived under the linear model \eqref{eq::lm}, and therefore, these two conflicting results seem paradoxical.

If we go back to the derivations above carefully, we will find that Section \ref{sec::vif} assumes that the $z_i$'s and $x_i$'s are both fixed, but Section \ref{sec::ancova} assumes that the $z_i$'s are random and the $x_i$'s are fixed. Therefore, the VIF and the ANCOVA results hold under different assumptions on the treatment indicators. This vaguely explains the paradox.

Technically, the settings for VIF and ANCOVA are slightly different, for example, $z_i$ can be general and may have arbitrary correlation with $x_i$  in the VIF result, but it is binary and arises from complete randomization in the ANCOVA result.  The data generating process below comes from the intersection of the settings for the two results, which allows for more unified discussion of them. 
 
Consider the following data generating process: for $i=1,\ldots, n$,
\begin{enumerate}
[(a)]
\item\label{eq::x}
the $x_i$'s are  fixed constants, centered at $\bar{x} = n^{-1} \sum_{i=1}^n x_i =0$;
\item\label{eq::y0}
generate the potential outcomes under control as
$y_i(0) = \alpha +  \beta' x_i + \varepsilon_i$, where $\mathcal{E} = (\varepsilon_1, \ldots, \varepsilon_n)$ are IID with mean $0$ and variance $\sigma^2$;
\item\label{eq::y1}
generate the potential outcomes under treatment as
$
y_i(1) =  y_i(0) + \tau ,
$
i.e., the individual treatment effect $y_i(1) - y_i(0)$ is constant $\tau$;
\item\label{eq::ZZ}
generate $Z = (z_1,\ldots, z_n)$   from a random permutation of $n_1$ 1's and $n_0$ 0's; 
\item\label{eq::obs}
the observed outcome is 
\begin{eqnarray}
y_i &=& z_iy_i(1) + (1-z_i)y_i(0)  \nonumber    \\
&=& \tau z_i  +  y_i(0) \nonumber    \\
&=& \alpha +  \tau z_i + \beta'x_i + \varepsilon_i .
\label{eq::lm2}
\end{eqnarray}
\end{enumerate}
In \eqref{eq::y0} and \eqref{eq::y1}, I use the potential outcomes notation \citep{Neyman23}. Readers who are uncomfortable with $y_i(1)$ and $y_i(0)$ can ignore steps \eqref{eq::y0} and \eqref{eq::y1} and view \eqref{eq::lm2} as the data generating process with random $\varepsilon_i$'s and $z_i$'s. Then $\tau$ represents the average treatment effect, which is the parameter of interest.

\subsection{Comparing the variances} 

Conditional on $Z $, \eqref{eq::lm2} is a linear model with fixed $(z_i, x_i)$'s and homoskedastic errors $\varepsilon_i$'s. The discussion in Section \ref{sec::vif} applies in this case. Then from the VIF result, we know that 
\begin{equation}
\var(\hat{\tau}_\text{a} \mid Z) \geq \var(\hat{\tau} \mid Z), 
\label{eq::conditionalvariances}
\end{equation}
i.e., the estimator adjusting for covariates $x_i$'s has larger variance. However, $\hat{\tau}_\text{a}$ is unbiased but $ \hat{\tau} $ is biased. From the classic OLS theory, 
$$
E( \hat{\tau}_\text{a} \mid Z) = \tau,
$$ 
and from the formula of \eqref{eq::adj-2}, the bias of $\hat{\tau}$ is 
$$
E( \hat{\tau} \mid Z) - \tau = \beta '  \hat{\delta}_x. 
$$ 
Therefore, the smaller conditional variance of $\hat{\tau}$ comes at the cost of having a larger conditional bias.  The conditional bias of $\hat{\tau} $ vanishes only in the special case with $\hat{\delta}_x=0$, that is, the covariates are perfectly balanced in means across the treatment and control groups.

Averaging over $Z$, we have random potential outcomes and random treatment indicators. The discussion in Section \ref{sec::ancova} applies in this case. We have shown that 
$$
E( \hat{\tau}   ) = \tau,\quad 
E( \hat{\tau}_\text{a}   ) \approx \tau,
$$ and moreover, asymptotically (ignoring higher order terms), 
\begin{equation}
\var( \hat{\tau}  ) \geq \var( \hat{\tau}_\text{a}   ) .
\label{eq::unconditionalvariances}
\end{equation}

Mathematically, the efficiency reversal results in \eqref{eq::conditionalvariances} and \eqref{eq::unconditionalvariances} do not lead to contradiction given the explicitly specified conditioning sets. Statistically, however, they form a paradox that is similar to the classic Simpson's paradox of effect reversal due to different conditioning sets. This paradox can be explained by decompositions based on the law of total variance:
\begin{eqnarray*}
\var(\hat{\tau}_\text{a}) &=&  E \{ \var(\hat{\tau}_\text{a} \mid Z)  \}
+ \var \{ E(\hat{\tau}_\text{a} \mid Z)  \}  \\
&=& E \{ \var(\hat{\tau}_\text{a} \mid Z)  \} + \var(\tau) \\
&=&  E \{ \var(\hat{\tau}_\text{a} \mid Z)  \} 
\end{eqnarray*}
and 
\begin{eqnarray*}
\var(\hat{\tau}) &= & E \{ \var(\hat{\tau} \mid Z)  \}
+ \var \{ E(\hat{\tau} \mid Z)  \} \\
&=& E \{ \var(\hat{\tau} \mid Z)  \} + \var( \tau + \beta '  \hat{\delta}_x ) \\
&= & E \{ \var(\hat{\tau} \mid Z)  \} + \var(   \beta '  \hat{\delta}_x ).
\end{eqnarray*}
Based on the VIF result, $E \{ \var(\hat{\tau}_\text{a} \mid Z)  \} \geq E \{ \var(\hat{\tau} \mid Z)  \}$, but their difference is small because the $R^2_{z\mid x}$ between $z_i$ and $x_i$ is close to zero under complete randomization. We can ignore their difference in asymptotic analysis. More importantly,  the unadjusted estimator has an additional variance term due to the conditonal bias which reverses the ordering of $\var(\hat{\tau}_\text{a}) $ and $\var(\hat{\tau}) $.

\subsection{Comparing the estimated variances}\label{subse::lukemiratrix}

Sections \ref{sec::vif} and \ref{sec::ancova} compare the variances of $\hat{\tau}_\text{a} $ and $\hat{\tau}$ which are theoretical quantities depending on the unknown true data generating process. In practice, standard statistical software packages report the estimated variances based on OLS:
$$
\hat{\var}(\hat{\tau}_a) = \frac{ \hat{\sigma}^2_{y\mid z,x}}{  \sumn (z_i - \bar{z}) ^2 } \times \frac{1}{1-R^2_{z\mid x}}
$$
and
$$
\hat{\var}(\hat{\tau}) = \frac{ \hat{\sigma}^2_{y\mid z}}{  \sumn (z_i - \bar{z}) ^2 } ,
$$
where $ \hat{\sigma}^2_{y\mid z,x} $ equals the residual sum of squares divided by  $n-2-\text{dim}(x)$ in the OLS fit of $y_i$ on $(1,z_i,x_i)$, and $\hat{\sigma}^2_{y\mid z}$ equals the residual sum of squares divided by $n-2$ in the OLS fit of $y_i$ on $(1,z_i)$. The ratio of these two variances depends on $ \hat{\sigma}^2_{y\mid z,x} / \hat{\sigma}^2_{y\mid z}$ and $R^2_{z\mid x}$, which can be larger or smaller than $1$. Importantly, this is a numeric result regardless of whether or not we condition on $Z$. 

In fact, under the data generating process in Section \ref{subsec::dgp}, $\hat{\var}(\hat{\tau}_a) $ is often smaller than $\hat{\var}(\hat{\tau}) $ as long as the covariates are predictive to the outcome. This is true due to two basic facts: first, $R^2_{z\mid x}$ is close to $0$ under complete randomization so the VIF can be ignored asymptotically; second, $ \hat{\sigma}^2_{y\mid z,x} $ is often smaller than $\hat{\sigma}^2_{y\mid z}$ because the residual sum of squares decreases with an additional predictive covariate. This argument ignores the opposite impact of the degrees of freedom correction, which is reasonable when the sample size is large and the dimension of covariates is small. See \citet{freedman1983note} for discussion of $R^2$ with high dimensional covariates.

The above heuristic comparison of $\hat{\var}(\hat{\tau}_a)$ and $\hat{\var}(\hat{\tau}) $  is not in contradiction with the VIF result which concerns the true variances conditional on $Z$. The estimated variances can be different from the true variances, especially when the linear model of $y_i$ on $(1,z_i)$ is misspecified.

\section{Connection with randomization inference}

\subsection{Variances conditional on the error terms}
Another conditioning scheme leads to discussion beyond Sections \ref{sec::vif} and \ref{sec::ancova}. Conditional on the error terms $\mathcal{E}$, we have fixed potential outcomes and completely randomized $Z$. Statistics under this regime is called randomization inference, or design-based inference. The classic results from randomization inference are $E( \hat{\tau} \mid \mathcal{E}) = \tau$  \citep{Neyman23}, $E( \hat{\tau}_\text{a} \mid \mathcal{E}) \approx \tau$ \citep{freedman2008regression_b, lin2013agnostic}, and
$
\var( \hat{\tau} \mid \mathcal{E}  )  \geq \var( \hat{\tau}_\text{a} \mid \mathcal{E} )
$ 
asymptotically \citep{freedman2008regression_b, lin2013agnostic}. This relative efficiency is coherent with $\var(\hat{\tau})\geq   \var(\hat{\tau}_\text{a}) $ asymptotically. The coherence can be explained by decompositions  based on the law of total variance:
\begin{eqnarray*}
\var(\hat{\tau}_\text{a}) &=&  E \{ \var(\hat{\tau}_\text{a} \mid \mathcal{E})  \}
+ \var \{ E(\hat{\tau}_\text{a} \mid \mathcal{E})  \}  \\
&\approx & E \{ \var(\hat{\tau}_\text{a} \mid \mathcal{E})  \} + \var(\tau) \\
&=&  E \{ \var(\hat{\tau}_\text{a} \mid \mathcal{E})  \} 
\end{eqnarray*}
and 
\begin{eqnarray*}
\var(\hat{\tau}) &= & E \{ \var(\hat{\tau} \mid \mathcal{E})  \}
+ \var \{ E(\hat{\tau} \mid \mathcal{E})  \} \\
&=& E \{ \var(\hat{\tau} \mid \mathcal{E})  \} + \var( \tau   ) \\
&= & E \{ \var(\hat{\tau} \mid \mathcal{E})  \} ,
\end{eqnarray*}
where both $\hat{\tau}_\text{a}$ and $\hat{\tau}$ are unbiased for $\tau$ at least asymptotically. 
These results are all in favor of ANCOVA which improves the estimation efficiency.
I summarize the results in Table \ref{tb::summary}.

\begin{table}
\caption{Comparison under the data generating process in \eqref{eq::x}--\eqref{eq::obs}}\label{tb::summary}
\begin{tabular}{|c|c|c|c|}
\hline 
& mean of $\hat{\tau}_\text{a}$ & mean of $\hat{\tau}$ & variance comparison \\
\hline 
unconditional& $E( \hat{\tau}_\text{a}) = \tau$ & $E( \hat{\tau}) = \tau$& 
$\var( \hat{\tau}_\text{a}) \leq \var( \hat{\tau})$ asymptotically\\ \hline 
conditional on $Z$& $E( \hat{\tau}_\text{a}\mid Z) = \tau$ & $E( \hat{\tau}\mid Z) = \tau + \beta'\hat{\delta}_x$ &
$\var( \hat{\tau}_\text{a}\mid Z) \geq \var( \hat{\tau}\mid Z)$ \\\hline 
conditional on $\mathcal{E}$ & $E( \hat{\tau}_\text{a}\mid \mathcal{E}) \approx \tau$ & $E( \hat{\tau}\mid \mathcal{E}) = \tau$& 
$\var( \hat{\tau}_\text{a}\mid \mathcal{E})  \leq \var( \hat{\tau}\mid \mathcal{E})$ asymptotically\\
\hline 
\end{tabular}
\end{table}

\subsection{More general potential outcomes model}

The data generating process in \eqref{eq::x}--\eqref{eq::obs} assumes constant treatment effect and homoskedastic errors. It yields the standard ANCOVA model \eqref{eq::lm} or \eqref{eq::lm2}. The literature of randomization-based causal inference often deals with more general potential outcomes, without requiring these linear model assumptions \citep{Neyman23, freedman2008regression_b, lin2013agnostic, ImbensRubin15, li2017general}. In general cases with possibly mis-specified linear models, \citet{freedman2008regression_b} showed that ANCOVA may increase or decrease the efficiency compared to $\hat{\tau}$. \citet{lin2013agnostic} recently proposed to use a modified ANCOVA estimator that also includes the interaction term $z_i \times x_i$ in OLS, and showed that this estimator is at least as efficient as $\hat{\tau}$ asymptotically.

\subsection{A design issue} 
 
From the above discussion, $\hat{\delta}_x$ is the key quantity that causes the paradox. If it is zero, then $\hat{\tau}_\text{a} = \hat{\tau}$ and the paradox disappears. Complete randomization ensures $E( \hat{\delta}_x ) = 0$, but in a particular allocation, $\hat{\delta}_x$ can differ from zero.  From the experimental design perspective, $\hat{\delta}_x$ measures the covariate balance across treatment and control groups. Complete randomization ensures covariate balance on average, but a particular allocation may have covariate imbalance. \citet{morgan2012rerandomization} proposed to use rerandomization to improve the data generating process \eqref{eq::ZZ} by forcing the treatment indicators $Z$ to satisfy 
$$ 
\hat{\delta}_x '  \cov(\hat{\delta}_x )^{-1}  \hat{\delta}_x  =
\hat{\delta}_x '   \left(   \frac{n}{n_1n_0} S_x^2   \right)^{-1}    \hat{\delta}_x  \leq  a,
$$
where $a >0$ is a predetermined threshold.  Under the randomization inference framework, \citet{li2018asymptotic} show that  this new experimental design improves the efficiency of  $\hat{\tau}$, which is, in fact, close to the efficiency of $\hat{\tau}_\text{a}$ for small $a \approx 0$. From our discussion before, rerandomization can also reduce the conditional bias of   $\hat{\tau}$ given $Z$ because it forces $\hat{\delta}_x$ to be small for any realized value of $Z.$ Therefore, rerandomization can mitigate the paradox through experimental design.  See \citet{li2019rerandomization} for more unified discussion.

\section{Final remarks}

I have shown that the seemingly paradoxical results of VIF and ANCOVA are due to different statistical assumptions. The key difference is whether or not statistical inference is conditional on the treatment indicators $Z$. Conditioning on $Z$, the unadjusted estimator has smaller variance but  larger bias. Averaging over  $Z$, both unadjusted and adjusted estimators are consistent for $\tau$ but the variance of the adjusted estimator is no larger than that of the unadjusted estimator. In randomized experiments, we recommend using ANCOVA under a constant treatment effect model or its modified version for general settings \citep{lin2013agnostic, li2019rerandomization}. 

I end this note with two minor technical issues.  First, I assume that the $x_i$'s are fixed throughout the paper. With random covariates, we can  condition on them to obtain the same results. The key in the discussion is whether or not to condition on $Z$. 
Second, if the $Z_i$'s are  IID Bernoulli random variables as in a Bernoulli experiment, we can condition on $(n_1, n_0)$ to reduce the discussion to complete randomization.

\section*{Acknowledgments}

The author thanks Ugur Yildirim for raising this question in his class of ``Linear Models'' at University of California, Berkeley, Luke Miratrix for the discussion of estimated variances in Section \ref{subse::lukemiratrix}, and Anqi Zhao, Xinran Li, Liyun Chen, Zhichao Jiang, and Yuting Ye for helpful discussions. A reviewer made many constructive comments that improved the paper significantly. This research was partially support by the U. S. National Science Foundation (grant \# 1945136).

\bibliographystyle{apalike}
\bibliography{causal}

\end{document}